\documentclass[graybox]{svmult}

\usepackage{mathptmx}       
\usepackage{helvet}         
\usepackage{courier}        
\usepackage{type1cm}        
\usepackage{makeidx}         
\usepackage{graphicx}        
\usepackage{multicol}        
\usepackage[bottom]{footmisc}

\usepackage{amssymb,amsfonts}
\usepackage{cite}

\newcommand{\be}[0]{\begin{equation}}
\newcommand{\ee}[0]{\end{equation}}

\newcommand{\F}{{\cal F}}
\newcommand{\N}{{\cal N}}

\newcommand{\Z}{\mathbb{Z}}
\newcommand{\Ka}{K{\"a}hler }
\renewcommand{\Re}{{\rm Re}\,}
\renewcommand{\Im}{{\rm Im}\,}

\newcommand{\abs}{|}

\newcommand{\ie}{{\em i.e.} }
\newcommand{\via}{{\it via} }

\newcommand{\where}{\mbox{where}}

\newcommand{\when}{\mbox{when}}
\renewcommand{\and}{\mbox{and}}

\begin{document}

\title*{Large volume supersymmetry breaking without decompactification problem}
\author{Herv\'e Partouche}
\institute{Herv\'e Partouche \at Centre de Physique Th\'eorique, Ecole Polytechnique, CNRS, Universit\'e Paris-Saclay\\
F--91128 Palaiseau cedex, France, \email{herve.partouche@polytechnique.edu}
}
\maketitle

\vspace{-2cm}
\abstract{We consider heterotic string backgrounds in four-dimensional Minkowski space, where $\N=1$ supersymmetry is spontaneously broken at a low scale $m_{3/2}$ by a stringy Scherk-Schwarz mechanism. We review how the effective gauge couplings at 1-loop may evade the ``decompactification problem", namely the proportionality of the gauge threshold corrections, with the  large  volume of the compact space involved in the supersymmetry breaking.}


\section{Introduction}
\label{intro}

A sensible physical theory must at least meet two  requirements : Be realistic and analytically under control. The first point can be satisfied by considering string theory, which has the advantage to be, at present time, the only setup in which both gravitational and gauge interactions can be described consistently at the quantum level. In this review, we do not consider cosmological issues and thus analyze models defined classically in four-dimensional Minkowski space. The ``no-scale models" are particularly interesting since, by definition, they describe in supergravity or string theory classical backgrounds, in which supersymmetry is spontaneously broken at an arbitrary scale $m_{3/2}$  in flat space \cite{noscale}. In other words, even if supersymmetry is not explicit, the classical vacuum energy vanishes. 

The most conservative way to preserve analytical control is to ensure the validity of perturbation theory. In string theory, quantum loops can be evaluated explicitly, when the underlying two-dimensional conformal field theory is itself under control. Clearly, this is the case, when one considers free field on the world sheet, for instance in toroidal orbifold models \cite{orbifolds} or fermionic constructions \cite{fermionic}. In these frameworks, the $\N=1\to \N=0$ spontaneous breaking of supersymmetry can be implemented at tree level \via a stringy version \cite{SSstring} of the Scherk-Schwarz mechanism \cite{SS}.\footnote{Note that non-perturbative mechanisms based on gaugino condensation could also be considered, but only at the level of the low energy effective supergravity, thus at the price of loosing part of the string predictability.} In this case, the supersymmetry breaking scale is of order of the inverse volume of the internal directions involved in the breaking. For a single circle of radius $R$, one has
\be
m_{3/2}={M_{\rm s}\over R}\, ,
\ee
where $M_{\rm s}$ is the string scale, so that having a low $m_{3/2}=O(10\mbox{ TeV})$  imposes the circle to be extremely large, $R=O(10^{17})$ \cite{AntoniadisTeV}. Such large directions yield towers of light Kaluza-Klein states and a problem arises from those charged 
under some gauge group factor $G^i$. In general, their contributions to the quantum corrections to the inverse squared gauge coupling is proportional to the very large volume and invalidates the use of perturbation theory. 

To be specific, let us consider  in heterotic string the 1-loop low energy running gauge coupling $g_i(\mu)$ , which satisfies \cite{thresholds}
\begin{equation}
\label{running}
{16\pi^2\over g^2_i(\mu)}=k^i{16\pi^2\over g^2_{\rm s}}+b^i\ln {M^2_{\rm s}\over \mu^2}+\Delta^i\, .
\end{equation} 
In this expression, $g_{\rm s}$ is the string coupling and $k^i$ is the Kac-Moody level of $G^i$. The logarithmic contribution, which depends on the energy scale $\mu$, arises from the massless states and is proportional to the $\beta$-function coefficient $b^i$, while the massive modes yield the threshold corrections $\Delta^i$. The main contributions to the latter arise from the light Kaluza-Klein states, which for a single large radius yield
\begin{equation}
\label{D}
\Delta^i=C^iR-b^i \ln R^2+O\left({1\over R}\right) ,
\end{equation}
where $C^i=Cb^i-C'k^i$, for some non-negative $C$ and $C'$ that depend on other moduli. When $C^i=O(1)$, requiring in Eq. (\ref{running}) the loop correction to be small compared to the tree level contribution imposes $g_{\rm s}^2 R<1$. In other words, for perturbation theory to be valid, the string coupling  must  be extremely weak, $g_{\rm s}<O(10^{-6.5})$. If $C^i>0$, which implies  $G^i$ is not asymptotically free, Eq. (\ref{running}) imposes the running gauge coupling to be  essentially free, $g_i(\mu)=O(g_{\rm s})$, and $G^i$ describes a hidden gauge group. However, if $C^i<0$, which is the case if $G^i$ is asymptotically free, the very large tree level contribution proportional to $1/g_{\rm s}^2$ must cancel $C^i R$, up to very high accuracy, for the running gauge coupling to be of order 1 and have a chance to describe realistic gauge interactions.  This unnatural fine-tuning is a manifestation of the so-called ``decompactification problem'', which actually arises generically, when a  submanifold of the internal space is large, compared to the string scale, \ie when the internal conformal field theory  allows a geometrical interpretation in terms of a compactified space. 

To avoid the above described behavior, $C^i$ can be required to vanish. This is trivially the case in the $\N=4$ supersymmetric theories, where actually $b^i=0$ and $\Delta^i=0$. The condition $C^i=0$ remains valid in the theories realizing the $\N=4\to \N=2$ spontaneous breaking, provided $\N=4$ is recovered when the volume is sent to infinity \cite{solving}. In this case, the threshold corrections scale logarithmically with the volume and no fine-tuning is required for perturbation theory to be valid. In Sect. \ref{mod}, we review the construction of models that realize an $\N=1\to \N=0$ spontaneous breaking at a low scale $m_{3/2}$, while avoiding the decompactification problem. The corresponding threshold corrections are computed in Sect. \ref{blocks} \cite{N=0thresh, planck15}. 
 

\section{The non-supersymmetric $\Z_2\times \Z_2$ models}
\label{mod}

In the present work, we focus on heterotic string backgrounds in four-dimensional Minkowski space and analyze the gauge coupling threshold corrections. At 1-loop, their formal expression is \cite{thresholds,universality,FlorakisN=0}
\begin{eqnarray}
\Delta^i&=&\int_{\cal F}{d^2\tau \over \tau_2}\!
\left(
{1\over 2}\sum_{a,b}{\cal Q}\!\big[{}^a_b\big]\!(2v)
\left({\cal P}_i^2(2\bar w) - {k^{i}\over 4\pi{\tau_2}}\right)\! \tau_2 \, Z\!\big[{}^a_b\big]\!(2v,2\bar w)-b^i
\right)\!\!\Bigg\abs_{v=\bar w=0}
\nonumber\\
&&+b^i\log {2\,e^{1-\gamma}\over \pi\sqrt{27}} \, ,
\label{2}
\end{eqnarray}
where $\F$ is the fundamental domain of $SL(2,\Z)$ and $Z\!\big[{}^a_b\big]\!(2v,2\bar w)$ is a refined partition function for given spin structure $(a,b)\in\Z_2\times \Z_2$.  ${\cal P}_i(2\bar w)$ acts on the right-moving sector as the squared charge operator of the gauge group factor $G^i$, while ${\cal Q}\!\big[{}^a_b\big]\!(2v)$ acts on the left-moving sector as the helicity operator,\footnote{Our conventions for the Jacobi functions $\theta\!\big[{}^a_b\big]\!(\nu\abs\tau)$ (or $\theta_{\alpha}(\nu\abs\tau)$, $\alpha=1,\dots,4$) and Dedekind function can be found in \cite{KiritsisBook}.} 
\begin{equation}
\label{Qop}
{\cal Q}\!\big[{}^a_b\big]\!(2v) = {1\over 16\pi^2}\, {\partial_v^2(\theta\!\big[{}^a_b\big]\!(2v))\over \theta\!\big[{}^a_b\big]\!(2v)}-{i\over\pi}\, \partial_\tau\log\eta\equiv {i\over \pi}\, \partial_{\tau}\!\left(\log {\theta\!\big[{}^a_b\big]\!(2v)\over\eta} \right) .
\end{equation}

From now on, we consider  $\Z_2\times \Z_2$ orbifold models \cite{orbifolds} or fermionic constructions \cite{fermionic} in which the marginal deformations parameterized by the \Ka and complex structures $T_I,U_I$, $I=1,2,3$, associated to the three internal 2-tori are switched on \cite{modulidef,N=0thresh}. In both cases, orbifolds or ``moduli-deformed fermionic constructions", $\N=1$ supersymmetry is spontaneously broken by a stringy Scherk-Schwarz mechanism \cite{SSstring}. 
The associated genus-1 refined partition function is
\begin{eqnarray}
\label{Z}
   Z(2v,2\bar w)=&&\!\!\!{1\over \tau_2(\eta \bar\eta)^2}\times  \\
   &&\!\!\!{1\over 2} \sum_{a,b}\,  {1\over 2}  \sum_{H_1,G_1} \, {1\over 2}  
\sum_{H_2,G_2} (-1)^{a+b+ab}\, {\theta\!\big[{}^a_b\big]\!(2v) \over \eta}\, 
{\theta\!\big[{}^{a+H_1}_{b+G_1}\big]\over \eta}\,{\theta\!\big[{}^{a+H_2}_{b+G_2}\big]\over \eta}\, {\theta\!\big[{}^{a+H_3}_{b+G_3}\big] \over \eta}\times \nonumber\\
&&\!\!\!{1\over 2^N}\sum_{h_I^i,g_I^i}S_L\Big[{}^{a,\, h^i_I,\,  H_I}_{b,\,  g^i_I,\, G_I}  \Big]  \, Z_{2,2}\Big[{}^{h_1^i}_{g_1^i} 
 \Big \abs {}^{H_1}_{G_1} \Big]\, 
 Z_{2,2}\Big[{}^{h_2^i}_{g_2^i} \Big \abs{}^{H_2}_{G_2} \Big]\,
 Z_{2,2}\Big[{}^{h_3^i}_{g_3^i} \Big \abs{}^{H_3}_{G_3} \Big] \,
 Z_{0,16}\Big[{}^{h_I^i , \,  H_I}_{g_I^i ,\, G_I}  \Big]\!(2\bar w)  ,\nonumber 
\label{partition}
\end{eqnarray}
where our notations are as follows :
\begin{itemize}

\item The $Z_{2,2}$ conformal blocks arise from the three internal 2-tori. The genus-1 surface having two non-trivial cycles, $(h^i_I, g^i_I)\in \Z_2\times \Z_2$, $i=1,2$, $I=1,2,3$ denote associated shifts of the six coordinates. Similarly, $(H_I,G_I)\in \Z_2\times \Z_2$ refer to the twists, where we have defined for convenience $(H_3,G_3)\equiv (-H_1-H_2,-G_1-G_2)$. Explicitly, we have 

\be
Z_{2,2}\Big[{}^{h_I^1,\, h_I^2} _{g_I^1,\, g_I^2} \Big \abs {}^{H_I}_{G_I} \Big]\!=\left\{
\begin{array}{ll}
\displaystyle {\Gamma_{2,2}\Big[{}^{h_I^1,\, h_I^2}_{g_I^1,\, g_I^2} \Big]\!(T_I,U_I)\over (\eta\bar\eta)^2}\, ,\phantom{\delta_{\huge\abs{}^{h_I^1\;\;H_I}_{g_I^1\;\; G_I}\big\abs,0}}\mbox{ when }  (H_I, G_I)=(0,0)\mbox{ mod 2}\, ,\\ 
\displaystyle {4\eta \bar \eta\over\ \theta\!\big[{}^{1-H_I}_{1-G_I}\big]\, \bar\theta\!\big[{}^{1-H_I}_{1-G_I}\big]}\; 
\delta_{\Big\abs{}^{h_I^1\;\;H_I}_{g_I^1\;\; G_I}\Big\abs,0 \,\mbox{\scriptsize mod} \,2}\;\delta_{\Big\abs{}^{h_I^2\;\;H_I}_{g_I^2\;\; G_I}\Big\abs,0 \,\mbox{\scriptsize mod} \,2}\qquad \,  \mbox{otherwise}\, ,
\end{array}
\right.
\ee
where $\Gamma_{2,2}$ is a shifted lattice that depends on the  \Ka and complex structure moduli $T_I,U_I$ of the $I^{\rm th}$ 2-torus. The arguments of the Kronecker symbols are determinants. 

\item When defining each model, linear constraints on the shifts $(h^i_I,g^i_I)$ and twists $(H_I,G_I)$ may be imposed, leaving effectively $N$ independent shifts. 

\item $Z_{0,16}$ denotes the contribution of the 32 extra right-moving world sheet fermions. Its dependance on the shifts and twists may generate discrete Wilson lines, which break partially $E_8\times E_8$ or $SO(32)$. 

\item The first line contains the contribution of the spacetime  light-cone bosons, while the second is that of the left-moving  fermions.
  
\item $S_L$ is a conformal block-dependent sign that implements the stringy Scherk-Schwarz mechanism. A choice of $S_L$ that correlates the spin structure $(a,b)$ to some shift $(h^i_I,g^i_I)$ implements the $\N=1\to\N=0$ spontaneous breaking.  

\end{itemize}

The $\Z_2\times \Z_2$ models contain three $\N=2$ sectors. For the decompactification problem not to arise, we impose one of them to be realized as a spontaneously broken phase of $\N=4$. This can be done by demanding the $\Z_2$ action characterized by $(H_2,G_2)$ to be free. The associated generator twists the $2^{\rm nd}$ and $3^{\rm rd}$ 2-tori (\ie the directions $X^6,X^7,X^8,X^9$ in bosonic language) and shifts some direction(s) of the $1^{\rm st}$ 2-torus, say $X^5$ only. To simplify our discussion, we take the generator of the other $\Z_2$, whose action is characterized by $(H_1,G_1)$, to not be free : It twists the $1^{\rm st}$ and $3^{\rm rd}$ 2-tori, and fixes the $2^{\rm nd}$ one. Similarly, we suppose that the product of the two generators, whose action is characterized by $(H_3,G_3)$, twists the $1^{\rm st}$ and $2^{\rm nd}$ 2-tori, and fixes the $3^{\rm rd}$ one. These restrictions impose the moduli $T_2,U_2$ and $T_3,U_3$ not to be far from $1$, in order to avoid  the decompactification problem  to occur from the remaining two $\N=2$ sectors. However, our care in choosing the orbifold action is allowing us to take the volume of the $1^{\rm st}$ 2-torus to be large. 

The above remarks have an important consequence, since the final stringy Scherk-Schwarz mechanism responsible of the $\N=1\to \N=0$ spontaneous breaking must involve the moduli $T_1,U_1$ only, for the gravitino mass to be light. Thus, this breaking must be implemented \via a shift along the $1^{\rm st}$ 2-torus, say $X^4$, and a non-trivial choice of $S_L$. Therefore, the sector $(H_1,G_1)=(0,0)$  realizes the pattern of spontaneous breaking $\N=4\to \N=2\to \N=0$, while the  other two $\N=2$ sectors, which have $2^{\rm nd}$ and $3^{\rm rd}$ 2-tori respectively fixed, are independent of $T_1$ and $U_1$ and thus  remain supersymmetric. As a result, we have in the two following independent modular orbits :
\begin{eqnarray}
\label{S}
&S_L=(-1)^{ag^1_1+bh^1_1+h^1_1g^1_1}\,,&\when\quad (H_1,G_1)=(0,0)\, ,\nonumber \\
&\!\!\!\!\!\!\!\!\!\!\!\!\!\!\!\!\!\!\!\!\!\!\!\!\!\!\!\!\!\!\!\!\!\!\!\!\!\!\!\!\!\!\!\!\!\!\!\!\!\!\!\!\!\!\!\!\!S_L=1\, ,&\when \quad(H_1,G_1)\neq(0,0)\, .
\end{eqnarray}

Given the fact that we have imposed $(h^2_1,g^2_1)\equiv (H_2,G_2)$, the $1^{\rm st}$ 2-torus lattice takes the explicit form
\begin{eqnarray}
&\displaystyle\Gamma_{2,2}\Big[{}^{h_1^1,\, H_2} _{g^1_1,\, G_2} \Big]\!(T_1,U_1) =\sum_{m^i,n^i}&\!\!\!(-1)^{m^1g^1_1+m^2G_2}\, \displaystyle e^{2i\pi\bar\tau\left[m^1\left(n^1+{1\over 2}{h^1_1}\right)+m^2\left(n^2+{1\over 2}{H_2}\right)\right]}\times\nonumber\\
&&e^{-{\pi\tau_2\over \Im T_1\Im  U_1}\left\abs T_1\left(n^1+{1\over 2}{h^1_1}\right)+T_1U_1\left(n^2+{1\over 2}{H_2}\right)+U_1m^1-m^2\right\abs^2}.
\end{eqnarray}
This expression can be used to find the squared scales of spontaneous $\N=4\to \N=2$ and $\N=2\to \N=0$ breaking. For $\Re(U_1)\in (-{1\over 2},{1\over 2}]$, they are
\begin{equation}
\label{m32}
{M^2_{\rm s}\over \Im T_1\, \Im U_1}\; , \qquad m^2_{3/2}={\abs U_1\abs^2 M^2_{\rm s}\over \Im T_1\, \Im U_1}\, ,
\end{equation}
where the latter is nothing but the gravitino mass squared of the full $\N=0$ theory. For these scales to be small compared to $M_{\rm s}$, we consider the regime $\Im T_1\gg 1$, $U_1=O(i)$.


\section{Threshold corrections}
\label{blocks}

The threshold corrections can be evaluated in each conformal block \cite{N=0thresh}. Starting with those where $(H_1,G_1)=(0,0)$, the discussion is facilitated by summing over the spin structures. Focussing on the relevant parts of the refined partition fonction $Z$, we have
\begin{eqnarray}
&\displaystyle {1\over 2}\sum_{a,b}(-1)^{a+b+ab}& \!\!(-1)^{ag^1_1+bh^1_1+h^1_1g^1_1}\,\theta\!\big[{}^a_b\big]\!(2v)\, \theta \!\big[{}^a_b\big] \theta\!\big[{}^{a+H_2}_{b+G_2}\big] \theta\!\big[{}^{a-H_2}_{b-G_2}\big]= \nonumber \\
&&(-1)^{h^1_1g^1_1+G_2(1+h^1_1+H_2)}\, \theta\!\Big[{}^{1-h^1_1}_{1-g^1_1}\Big]^2(v) \,  \theta\!\Big[{}^{1-h^1_1+H_2}_{1-g^1_1+G_2}\Big]^2(v)  \,  ,
\label{theta}
\end{eqnarray}
which shows how many odd $\theta_1(v)\equiv \theta[^1_1](v)$ functions (or equivalently how many fermionic zero modes in the path integral) arise for given shift $(h^1_1,g^1_1)$ and twist $(H_2,G_2)$.\\

\vspace{0.5cm}
\noindent {\large \em Conformal block $A$ : \rm $(h^1_1,g^1_1)=(0,0)$, $(H_2,G_2)=(0,0)$} 

\noindent This block  is  proportional to $\theta\!\big[{}^1_1\big]^4(v)=O(v^4)$. Up to an overall factor $1/2^3$, it is the contribution of the $\N=4$ spectrum of the parent theory, when neither the $\Z_2\times \Z_2$ action nor the stringy Scherk-Schwarz mechanism are implemented. Therefore, it does not contribute to the 1-loop gauge couplings. \\

\noindent {\large \em Conformal blocks $B$ : \rm $(h^1_1,g^1_1)\neq (0,0)$, $(H_2,G_2)= (0,0)$} 

\noindent They are proportional to $\theta\!\Big[{}^{1-h^1_1}_{1-g^1_1}\Big]^4(v)=O(1)$. The parity of the winding number along the compact direction $X^4$ being $h^1_1$, the blocks with $h^1_1=1$ involve states, which are super massive compared to the pure Kaluza-Klein modes. These blocks are therefore exponentially suppressed, compared to the block  
$(h^1_1,g^1_1)=(0,1)$. Up to an overall factor $1/2^2$, the latter arises from the spectrum considered in the conformal block $A$, but in the $\N=4\to \N=0$ spontaneously broken phase, and contributes to the gauge couplings.\\

\noindent {\large \em Conformal blocks $C$ : \rm $(h^1_1,g^1_1)=(0,0)$, $(H_2,G_2)\neq (0,0)$} 

\noindent They are proportional to $\theta\!\big[{}^1_1\big]\!(v)^2\theta\!\big[{}^{1-H_2}_{1-G_2}\big]^2(v)=O(v^2)$ and do contribute to $\Delta^i$, due to the action of the helicity operator. Reasoning as in the previous case, the parity of the winding number along the compact direction $X^5$ is $H_2$, which implies the blocks with $H_2=1$ yield exponentially suppressed contributions, compared to that associated to the block $(H_2,G_2)=(0,1)$. Up to an overall factor $1/2^2$, the latter arises from a spectrum realizing the spontaneous $\N=4\to\N_C=2$ breaking, which contributes to the couplings. \\

\noindent {\large \em Conformal blocks $D$ : \rm $(h^1_1,g^1_1)=(H_2,G_2)\neq (0,0)$} 

\noindent They are proportional to $\theta\!\big[{}^{1-H_2}_{1-G_2}\big]^2(v)\theta\!\big[{}^1_1\big](v)^2=O(v^2)$. The situation is identical to that of the conformal blocks $C$, except that the generator of the $\Z_2$ free action responsible of the partial spontaneous breaking of $\N=4$ twists $X^6,X^7,X^8,X^9$ and shifts $X^4,X^5$. The dominant contribution to the threshold corrections arises again from the block $(H_2,G_2)=(0,1)$, which describes a spectrum  realizing the spontaneous $\N=4\to\N_D=2$ breaking. \\

\noindent {\large \em Conformal blocks $E$ : \rm $\Big\abs{}^{h^1_1\;\,H_2}_{g^1_1\; \,G_2}\Big\abs\neq0$} 

\noindent The remaining conformal blocks have non-trivial determinant $\Big\abs{}^{h^1_1\;\,H_2}_{g^1_1\; \,G_2}\Big\abs$, which implies $ \theta\!\Big[{}^{1-h^1_1}_{1-g^1_1}\Big]^2(v) \,  \theta\!\Big[{}^{1-h^1_1+H_2}_{1-g^1_1+G_2}\Big]^2(v)=O(1)$. However, this condition is also saying that $(h^1_1,H_2)\neq (0,0)$, which means the modes in these blocks have non-trivial winding number(s) along $X^4$, $X^5$ or both. Therefore, their contributions to the gauge couplings are non-trivial but exponentially suppressed. \\

Having analyzed all conformal blocks satisfying $(H_1,G_1)=(0,0)$, we  proceed with the study of the modular orbit $(H_1,G_1)\neq(0,0)$, where the sign $S_L$ is trivial. Since the $1^{\rm st}$ 2-torus is twisted, these blocks are independent of the moduli $T_1,U_1$ and thus $m_{3/2}$. They can be analyzed as in the case of $\Z_2\times\Z_2$, $\N=1$ supersymmetric models. Actually, summing over the spin structures, the relevant terms in the refined partition function $Z$ become 
\begin{eqnarray}
&\displaystyle {1\over 2}
\sum_{a,b}\!\!&(-1)^{a+b+ab}\theta\!\big[{}^a_b\big]\!(2v)\, \theta\!\big[{}^{a+H_1}_{b+G_1}\big] \theta\!\big[{}^{a+H_2}_{b+G_2}\big] \theta\!\big[{}^{a-H_1-H_2}_{b-G_1-G_2}\big]\!=\nonumber \\
&&(-1)^{(G_1+G_2)(1+H_1+H_2)}\, \theta\!\big[{}^{1}_{1}\big]\!(v) \, 
 \theta\!\big[{}^{1-H_1}_{1-G_1}\big]\!(v)\, \theta\!\big[{}^{1-H_2}_{1-G_2}\big]\!(v) \,  \theta\!\big[{}^{1+H_1+H_2}_{1+G_1+G_2}\big]\!(v) \, ,
\end{eqnarray}
which invites us to split the discussion in three parts. \\

\noindent {\large \em $\N=2$ conformal blocks, with fixed 2$^{nd}$ 2-torus : $(H_2,G_2)= (0,0)$} 

\noindent They are proportional to $\theta\!\big[{}^1_1\big]^2(v)\theta\!\big[{}^{1-H_1}_{1-G_1}\big]^2(v)=O(v^2)$. The $2^{\rm nd}$ internal 2-torus is fixed by the non-free action of the $\Z_2$ characterized by $(H_1,G_1)$. Adding the conformal block $A$, we obtain an $\N=2$ sector of the theory, up to an overall factor $1/2$ associated to the second $\Z_2$. This spectrum leads to non-trivial corrections to the gauge couplings.\\

\noindent {\large \em $\N=2$  conformal blocks, with fixed 3$^{rd}$ 2-torus : $(H_1,G_1)= (H_2,G_2)$} 

\noindent Thy are proportional to $\theta\!\big[{}^1_1\big]^2(v)\theta\!\big[{}^{1-H_1}_{1-G_1}]^2(v)=O(v^2)$. Actually, $(H_3,G_3)=(0,0)$, which means that the $3^{\rm rd}$ 2-torus is fixed by the combined action of the generators of the two $\Z_2$'s. Adding the conformal block $A$, one obtains the last  $\N=2$ sector of the theory, up to an overall factor $1/2$. Again, this spectrum yields a non-trivial contribution to the gauge couplings.\\

\noindent {\large \em $\N=1$ conformal blocks  : \rm $\big\abs{}^{H_1\;\,H_2}_{G_1\; \,G_2}\big\abs\neq0$} 

\noindent  The remaining blocks have non-trivial determinant, $\big\abs{}^{H_1\;H_2}_{G_1\; G_2}\big\abs\neq 0$, which implies they are proportional to $\theta\!\big[{}^{1}_{1}\big]\!(v) \, 
 \theta\!\big[{}^{1-H_1}_{1-G_1}\big]\!(v)\, \theta\!\big[{}^{1-H_2}_{1-G_2}\big]\!(v) \,  \theta\!\big[{}^{1+H_1+H_2}_{1+G_1+G_2}\big]\!(v) =O(v)$. Acting on them with the helicity operator, the result is proportional to 
\begin{eqnarray}
\partial_v^2\Big(\theta\!\big[{}^1_1\big]\!(v)\, \theta\!\big[{}^{1-H_1}_{1-G_1}\big]\!(v)\, \theta\!\big[{}^{1-H_2}_{1-G_2}\big]\!(v)&&\!\!\!\!\theta\!\big[{}^{1+H_1+H_2}_{1+G_1+G_2}\big]\!(v)\Big)\!\Big\abs_{v=0}\propto \nonumber \\
&&\partial_v^2\Big(\theta_1(v)\, \theta_2(v)\, \theta_3(v)\,  \theta_4(v)\Big)\!\Big\abs_{v=0}=0\, ,
\end{eqnarray}
thanks to the oddness of $\theta_1(v)$ and evenness of $\theta_{2,3,4}(v)$.
Thus, these conformal blocks do not  contribute to the thresholds.\\

In the class of models we consider, the effective running gauge coupling associated to some gauge groupe factor $G^i$ has a universal form at 1-loop \cite{N=0thresh}. It can be elegantly expressed in terms of three moduli-dependent squared mass scales arising from the corrections associated to the conformal blocks $B,C,D$, 
\be
M^2_B\!=\!{M^2_{\rm s}\over |\theta_2(U_1)|^4\, \Im T_1\, \Im U_1}\!,M^2_C\!=\!{M^2_{\rm s}\over |\theta_4(U_1)|^4\, \Im T_1\, \Im U_1}\!,M^2_D\!=\!{M^2_{\rm s}\over |\theta_3(U_1)|^4\, \Im T_1\, \Im U_1}\!,
\ee
which are of order $m^2_{3/2}$, and two more scales 
\be
M^2_I={M^2_{\rm s} \over 16\big\abs\eta(T_I)\abs^4 \, \big\abs\eta(U_I)\abs^4\, \Im T_I \, \Im U_I}\, , \quad I=2,3,
\ee
of order $M^2_{\rm s}$ that encode the contributions of the $\N=2$ sectors associated to the fixed $2^{\rm nd}$ and $3^{\rm rd}$ internal 2-tori. It is also useful to introduce a ``renormalized string coupling'' \cite{universality}, 
\begin{eqnarray}
{16\, \pi^2\over g_{\rm renor}^2}\!&=&\!{16\, \pi^2\over g_{\rm s}^2}-{1\over 2}Y(T_2,U_2)-{1\over 2}Y(T_3,U_3)\, , \nonumber \\
\where \quad  Y(T,U)\!&=&\!{1 \over 12}\int_{\cal F}{d^2\tau\over \tau_2}\,
\Gamma_{2,2}(T,U)\! \left[ \!\Big(\bar E_2-{3\over \pi\tau_2}\Big){\bar E_4 \bar E_6\over \bar\eta^{24}}-\bar j+1008 \right]\!,\;\; 
\end{eqnarray}
in which $\Gamma_{2,2}=\Gamma_{2,2}\big[{}^{0,\, 0}_{0,\, 0} \big]$ is the unshifted lattice, while for $q=e^{2i\pi\tau}$, $E_{2,4,6}=1+O(q)$ are holomorphic Eisenstein series of modular weights 2,4,6 and $j={1/q}+744+O(q)$ is holomorphic and modular invariant. 
The inverse squared 1-loop gauge coupling at energy scale $Q^2 =\mu^2 {\pi^2\over 4}$ is then
\begin{eqnarray}
\label{thfinal}
{16\, \pi^2\over g_{i}^2(Q)} = k^{i}{16\, \pi^2\over g_{\rm renor}^2}\!\!\! \!&&
\displaystyle -\, {b^i_{B}\over 4}\ln\left({Q^2\over Q^2+M^2_B}\right)
-{b^i_C\over 4}\ln\left({Q^2\over Q^2+M^2_C}\right)
-{b^i_{D}\over 4}\ln\left( {Q^2\over Q^2+M^2_D}\right)\nonumber \\
 && \displaystyle -\, {b^i_{2}\over 2}  \ln\left({Q^2\over M^2_{2}} \right) 
 -{b^i_{3}\over 2}\ln\left( {Q^2\over M^2_{3}}\right)+O\left({m^2_{3/2}\over M^2_{\rm s}}\right),
\end{eqnarray}
which depends only on five  model-dependent $\beta$-function coefficients and the Kac-Moody level. In this final result, we have shifted $M_{B,C,D}^2\to Q^2+M_{B,C,D}^2$ in order to implement the thresholds at which the sectors $B$, $C$ or $D$ decouple, \ie when $Q$ exceeds $M_B$, $M_C$ or $M_D$. Thus, this expression is valid as long as $Q$ is lower than the mass of the heavy states we have neglected the exponentially suppressed contributions \ie the string or GUT scale, depending on the model. Taking $Q$ lower than at least one of the scales $M_B$, $M_C$ or $M_D$, the r.h.s. of Eq. (\ref{thfinal}) scales as $\ln \Im T_1$, which is the logarithm of the large $1^{\rm st}$ 2-torus volume, as expected for the decompactification problem not to arise.  \\

To conclude, we would like to mention two important remarks. 
First of all, we stress that the $\Z_2\times \Z_2$ models, where a $\Z_2$ is freely acting and a stringy Scherk-Schwarz mechanism responsible of the final  breaking of $\N=1$ takes place, have non-chiral massless spectra. This is due to the fact that in the $\N=1$, $\Z_2\times \Z_2$ models, chiral families occur from twisted states localized at fixed points. In the models we have considered, fixed points localized on the $2^{\rm nd}$ and $3^{\rm rd}$ 2-tori can arise but are independent of the moduli $T_1,U_1$ \ie $m_{3/2}$. Thus, taking the large volume limit of the $1^{\rm st}$ 2-torus, where $\N=2$ supersymmetry is recovered, one concludes that the twisted states are actually hypermultiplets \ie couples of families and anti-families. 

Second, we point out that in the models analyzed in the present work, the conformal block $B$ is the only non-supersymmetric and non-negligible contribution to the partition function $Z$, and thus to the 1-loop effective potential. In Ref. \cite{snc,planck15}, it is shown that in some models, the latter is positive semi-definite. The motion of the moduli $T_2,U_2$ and $T_3,U_3$ is thus attracted to points \cite{stab spectators}, where the effective potential vanishes, allowing $m_{3/2}$ to be arbitrary. In other words, the defining properties of the no-scale models, namely arbitrariness of the supersymmetry breaking scale $m_{3/2}$ in flat space, which are valid at tree level, are extended to the 1-loop level. This very fact, characteristic of  the so-called ``super no-scale models", may have interesting consequences on the smallness of a cosmological constant generated at higher orders. In Ref. \cite{ADM}, other models having 1-loop vanishing cosmological constant are also considered, which however suffer from the decompactification problem. 


\begin{acknowledgement}
I am grateful to A. Faraggi and K. Kounnas, with whom the original work \cite{N=0thresh} has been done in collaboration. I would also like to thank C. Angelantonj, I. Antoniadis and J. Rizos for fruitful discussions, and the Laboratoire de Physique Th\'eorique of Ecole Normale Sup\'erieure for hospitality.
\end{acknowledgement}

\end{document}